\documentclass[doublecol]{epl2} 

\title{Strain effects on electronic structure of the iron selenide superconductor}

\author{M. J. Winiarski\inst{1} \and M. Samsel-Czeka\l a\inst{1} \and A. Ciechan\inst{2}}

\institute{                    
  \inst{1} Institute of Low Temperature and Structure Research, Polish Academy of Sciences, Ok\'olna 2, 50-422 Wroc\l aw, Poland\\
  \inst{2} Institute of Physics, Polish Academy of Sciences, al. Lotnik\'ow 32/46, 02-668 Warsaw, Poland
}

\abstract{The influence of {various strains on crystal and electronic structures of superconducting FeSe} has been studied {\it ab initio}. We consider changes in the Fermi surface nesting with a vector {$\mathbf{Q}=(0.5,0.5)\times(2\pi /a)$ as crucial for rising superconductivity (SC) mediated by spin-fluctuations (SF)}. Our results indicate that {the {\it c}-axis strained FeSe exhibits the most imperfect nesting, which enhances SF and, hence, also SC}. In turn, the {\it ab}-plane compressive strain {slightly weakens this} nesting while the tensile strain {destroys it completely}. These findings are consistent with reported earlier experimental dependencies of superconducting transition temperatures on strain in FeSe thin films.}

\begin{document}

\maketitle

\section{\label{sec1}Introduction}

Iron chalcogenides draw wide interest because of their simple structure and possible applications. The iron selenide, FeSe, has been found to be superconducting {below the transition temperature $T_c=$ 8 K} \cite{Hsu}.
{Although the pure FeTe is non-superconducting, $T_c$ is risen up to 15 K in the solid solutions FeSe$_{1-x}$Te$_x$ for $x=0.5$ \cite{Yeh, Fang, Mizuguchi}. However,} the highest $T_c$ of 37 K has been detected for bulk FeSe under hydrostatic pressure \cite{Mizuguchi-FeSe, Margadonna, Millican, Kumar, Okabe}. A {comparable} enhancement of $T_c$ {up to more than} 30 K was also reported for ternary compounds $A_x$Fe$_2$Se$_2$ with the alkali metals atoms {\it A} (= K, Rb, Cs) located between Fe-Se layers \cite{Guo, Fang-AFeSe, Luo, Ying}, which can be explained by a chemical pressure effect. Furthermore, superconductivity (SC) in a Cu-doped FeSe is restored by external pressure and at 7.8 GPa it reaches maximum $T_c$ {of 31 K} \cite{Schoop}.

{Also} the influence of non-hydrostatic strain on superconducting properties of FeSe has been investigated experimentally. {On one hand, in} lattice-mismatched epitaxial films {of} FeSe the tensile strain {onset} suppresses SC \cite{Nie}. {On the other hand, in similar films of} FeSe$_{0.5}$Te$_{0.5}$, the uniaxial ({\it c}-axis) strain increases $T_c$ \cite{Si}. The SC properties may be also tuned by the control parameter of the Se deficit in FeSe$_{1-x}$ \cite{Wang, Han, Li}.

In recent years, the electronic structure of superconducting iron chalcogenides has intensively been studied both theoretically {\cite{Subedi, Chen, Ciechan, Singh, Chadov}} and experimentally \cite{Kumar, Chen, Tamai, Nakayama, Miao}. The authors of these works have postulated that in FeSe {(and related 11-type systems)} the multi-gap nature of SC is connected with interband interactions between the holelike $\beta$ and electronlike $\delta$ Fermi surface (FS) sheets. In particular, {SC can be mediated by} antiferromagnetic spin fluctuations, {observed experimentally by e.g. NMR \cite{Imai}, which are} driven by the {imperfect} nesting with the $\mathbf{q}\sim(\pi,\pi)$ vector, spanning the above FS sheets in iron  chalcogenides {\cite{Subedi, Ciechan, Singh}}.

{To our knowledge, this is the first theoretical study of the influence of strain on the electronic structure of FeSe.} We focus {mainly} on both qualitative and quantitative description of the nested area between the $\beta$ and $\delta$ FS sheets of FeSe under various kinds of strain in the unit cell (u.c.). The relation between our findings and available experimental data of the $T_c$ {variations} in the strained FeSe compound is discussed.

\section{\label{sec2}Computational details}

Band structure calculations for FeSe have been carried out in the framework of the density functional theory (DFT). A full theoretical optimization of atomic positions and geometry of the tetragonal u.c. of the PbO-type ($P4/nmm$, No. 129) under various strains was performed with the Abinit package \cite{Abinit, PAW}, {using Projector Augmented Wave (PAW) pseudopotentials, generated with Atompaw software \cite{Atompaw}. Since FeSe is a metallic system without strong electron-electron correlations, for this purpose, only {the local density approximation (LDA)} \cite{JP} and generalized gradient approximation (GGA) \cite{GGA} were utilized}. The following cases of strain have been considered here: hydrostatic pressure, biaxial in the {\it ab}-plane compressive and tensile strains as well as uniaxial in the {\it c}-axis compressive strain. The valence-band basis of 3s3p3d;4s4p states were selected for both Fe and Se atoms. Based on these results, we employed the full potential local-orbital (
FPLO) band structure code \cite{FPLO}, in the scalar-relativistic mode, to compute {densities of states (DOSs) and} the Fermi surface changes. Since the FS nesting properties of FeSe are subtle, very dense {\bf k}-point meshes in the Brillouin zone (BZ) were used, {\it i.e.} 64$\times$64$\times$64 and 256$\times$256$\times$256 for the self-consistent field (SCF) cycle and FS plots, respectively. This extreme accuracy of the {\bf k}-point meshes allows for a detailed description of the investigated here FS features.

Finally, we determined numerically a nesting function: 

{$f_{nest}(\mathbf{q})=\Sigma_{\mathbf{k},n,n'}
\frac{[1-F_{n}^{\beta}({\mathbf{k}})]F_{n'}^{\delta}(\mathbf{k+q})}
{|E_{n}^{\beta}({\mathbf{k}})-E_{n'}^{\delta}(\mathbf{k+q})|}$, 

where $F_{n}^{\beta}$ and $F_{n'}^{\delta}$ are the Fermi-Dirac functions of states $n$ and $n'$ in bands $\beta$ and $\delta$, ($F=$ 0 or 1 for holes or electrons), respectively. In turn, $E_{n}^{\beta}$ and $E_{n'}^{\delta}$ are energy eigenvalues of these bands. The considered here $f_{nest}({\mathbf{q||Q}})$, were $\mathbf{Q}=(0.5,0.5)\times(2\pi /a)$ is the ideal nesting vector, reflects a frequency of an occurrence of a given vector $\mathbf{q}\sim(\pi,\pi)$ (having the same or different lenght from $\mathbf{Q}$) in the {\bf k}-space, spanning the FS sheets originating from the $\beta$ and $\delta$ bands.}

\section{\label{sec3} Results and discussion}

The changes of lattice parameters {\it a} and {\it c} in the tetragonal FeSe unit cell under hydrostatic pressure are displayed in fig. 1. This figure clearly shows that the LDA approach yielded reasonable results, being much better than those obtained by GGA (particularly for low pressure), compared with available experimental data of Ref. \cite{Kumar}. Hence, further only the LDA results will be presented. The variations of the parameters {\it a}, {\it c} and free {\it Z} position of selenium atoms ($Z_{Se}$) in tetragonal FeSe u.c. {calculated under hydrostatic pressure in comparison with those obtained for setting} both {\it ab}-plane and {\it c}-axis compressive strains are presented in {figs. 2. and 3}. As demonstrated in {fig. 2}, the {\it ab}-plane compressive strain increases the {\it c}/{\it a} ratio while the hydrostatic pressure and {\it ab}-plane tensile strain (not displayed here) makes the opposite effects. Moreover, the {\it c}-axis compression is 
followed by changes in the u.c. geometry showing a strong tendency to become cubic above 3.5 GPa. {Fig. 3} illustrates results of the corresponding relaxations of the atomic position $Z_{Se}$. This figure indicates that the largest increase of $Z_{Se}$ takes place under the {\it c}-axis strain. The revealed here distinct anisotropic behavior of the FeSe lattice parameters is connected with its layered character of the crystal structure along the {\it c}-axis, typical of iron chalcogenides. Our results show that the iron-selenium network in the {\it ab}-plane is more stable than that forming the interlayer connections of these planes along the {\it c}-axis. In general, these structural features have appeared to be characteristic of high temperature superconductors.

\begin{table}
\caption{Calculated lattice parameters {\it a}, {\it c/a} ratio, and free  atomic position, $z_{Se}$, in strained u.c. of FeSe.}
\label{table1}
\begin{tabular}{llll}
Strain (GPa) &  {\it a} (nm) & {\it c/a} & $Z_{Se}$\\ \hline
unstrained (0 GPa) & 0.3592 & 1.4988 & 0.2570 \\
hydrostatic (9 GPa) & 0.3515 & 1.4018 & 0.2836 \\
{\it ab}-plane compressive (3 GPa) & 0.3505 & 1.5666 & 0.2612 \\
{\it ab}-plane tensile (3 GPa) & 0.3703 & 1.4089 & 0.2529 \\
{\it c}-axis compressive (1 GPa) & 0.3631 & 1.3849 & 0.2680 \\
\end{tabular}
\end{table}

The parameters {\it a}, {\it c/a}, and $Z_{Se}$, chosen to determine DOS's and FS nesting properties (depicted in {figs. 4 and 6}) are collected in Table I. It is worth underlining that the presented parameters obtained under hydrostatic pressure of 9 GPa correspond to the experimental conditions for which $T_{c}$ of FeSe reaches its maximum value. For the other considered cases of applied strain, because of a lack of measured structural data, they have been chosen arbitrarily to detect any modifications of the electronic structure that may explain the changes of $T_{c}$ observed in such forms of FeSe as e.g. thin films \cite{Nie}.     

\begin{figure}
\onefigure[scale=0.8]{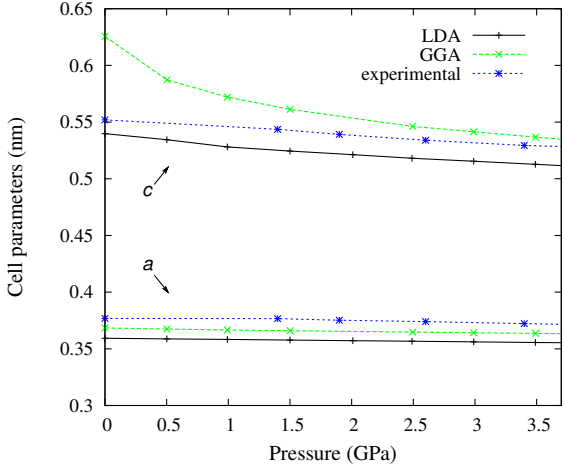}
\caption{{Variations of lattice parameters {\it a} and {\it c} of tetragonal FeSe u.c. with hydrostatic pressure, calculated by employing LDA and GGA approaches, compared with experimental data of Ref. \cite{Kumar}.}}
\label{Fig1} 
\end{figure}

\begin{figure}
\onefigure[scale=0.8]{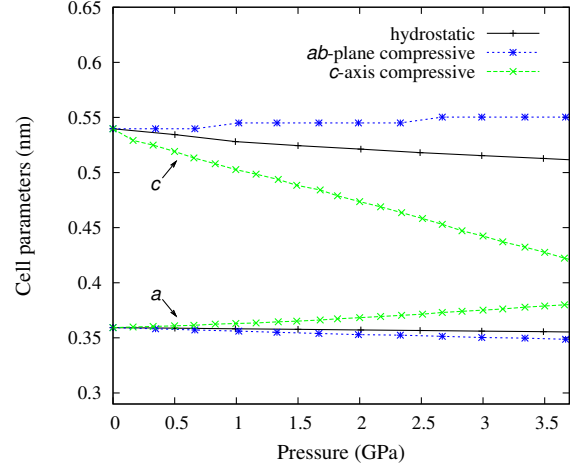}
\caption{Changes of lattice parameters {\it a} and {\it c} of tetragonal FeSe u.c. with different strain conditions.}
\label{Fig2} 
\end{figure}

\begin{figure}
\onefigure[scale=0.8]{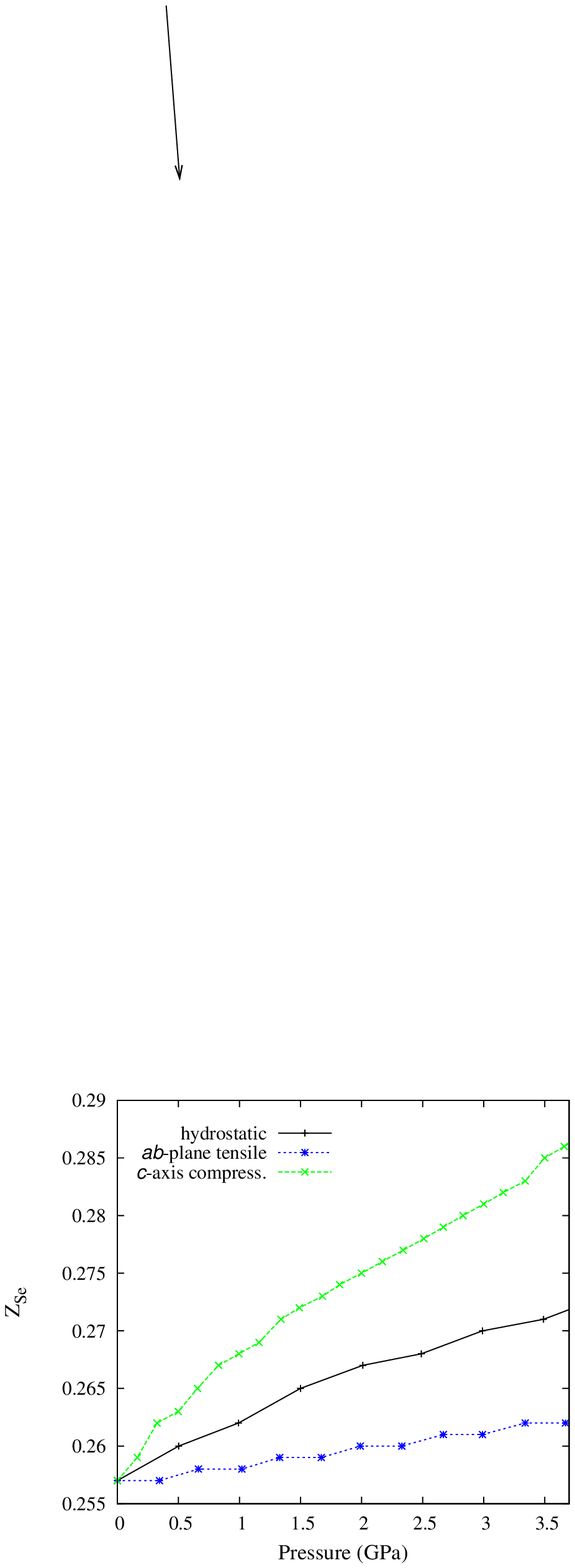}
\caption{The same as in {fig. 2} but for the free atomic $Z_{Se}$ position.}
\label{Fig3}
\end{figure}

As seen in {fig. 4}, any strain effects on the total DOS's at the Fermi level ($E_F$) seem to be irrelevant. However, it is worth noticing that both hydrostatic and {\it ab}-plane compressive strains make a shift of the whole DOS spectrum towards higher energies. In contrast, the {\it ab}-plane tensile and {\it c}-axis strains cause an opposite effect, which may lead to a rapid decrease of DOS at $E_F$ that, in turn, can suppress SC.  

{Since for FeSe, the {\it c}-axis strain has the most pronounced influence on the electronic structure, the corresponding FS's are presented in fig. 5.} It is seen in this figure that above 2 GPa, a topological transition in the FS takes place when both the cylindrical $\beta$ and $\delta$ sheets around the $\Gamma$ and $M$ points, respectively, are split into smaller FS pockets. This effect is much stronger in comparison with that of hydrostatic pressure reported previously \cite{Ciechan}. Therefore, in this work, we consider only small values of the {\it c}-axis strain to show an overall relation between the strain and electronic structure of FeSe.

\begin{figure}
\onefigure[scale=0.8]{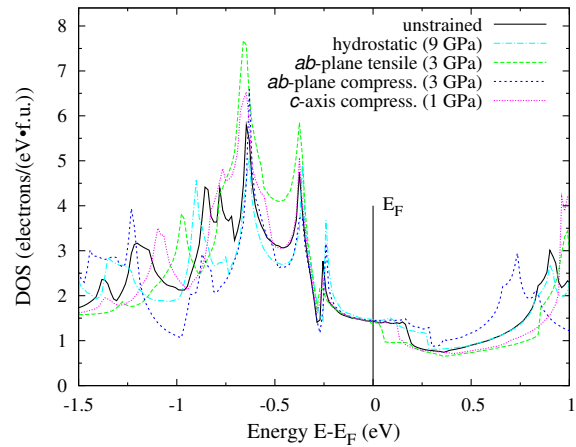}
\caption{Total DOS's plots of FeSe under various strain conditions.}
\label{Fig4}
\end{figure}

\begin{figure}
\onefigure[scale=0.5]{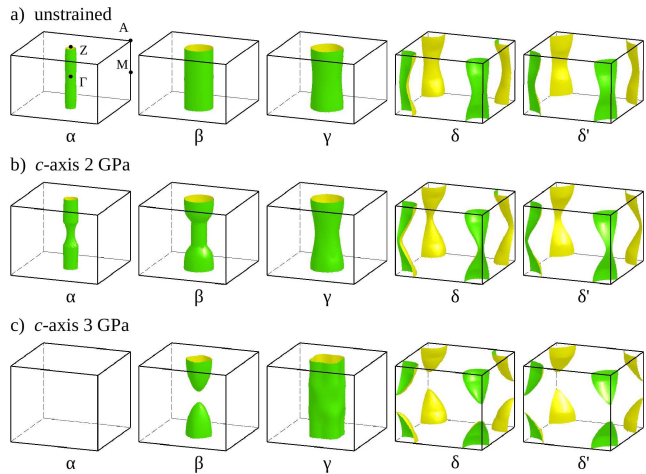}
\caption{Changes {in succeeding FeSe FS sheets (marked by Greek letters) under setting} the {\it c}-axis compressive strain.}
\label{Fig5} 
\end{figure}

\begin{figure}
\onefigure[scale=0.8]{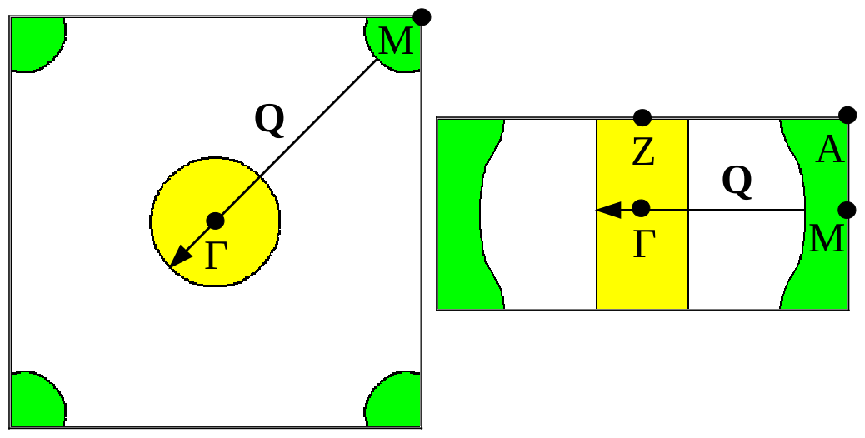}
\caption{The ideal nesting vector, {\textbf Q}, spanning the $\beta$ and $\delta$ FS sheets ({displayed in fig. 5(a)}), marked on two main FS sections of the unstrained FeSe.}
\label{Fig6}
\end{figure}

Next, we consider quantitatively the nesting properties of the $\beta$ and $\delta$ FS sheets along the direction of $\mathbf{Q}$, being parallel to the $\Gamma M$ and $ZA$ lines as visualised in {fig. 6}. The numerical nesting function, $f_{nest}$, determined for these sheets of FeSe (defined in Section II) {is presented in fig. 7}. {Part (a) of this figure} demonstrates that for the unstrained FS $f_{nest}$ reaches one pronounced maximum exactly at $\mathbf{Q}$, corresponding to the region in the vicinity of the $\Gamma M$ line, and the other maximum at $\mathbf{q_1}$ of a shorter length (0.625), dominating around the $ZA$ line. {As is shown in fig. 7(b),} the hydrostatic pressure diminishes the intensities of the peak at $\mathbf{Q}$ and $\mathbf{q_2}$ ({the latter} being shorter than $\mathbf{q_1}$ as well as the range of length of possible spanning vectors $\mathbf{q}$ becomes much wider than in the unstrained FeSe. This effect may {explain} an 
enhancement of spin fluctuations in FeSe. In unstrained FeSe, an antiferromagnetic spin ordering is stable {on account of} the high intensity peak at $\mathbf{Q}$\cite{Miao}. Thus, its instability, reflected by the smoothed $f_{nest}(\mathbf{Q})$ (imperfect nesting), should lead to spin fluctuations being responsible for SC.

\begin{figure}
\onefigure[scale=0.8]{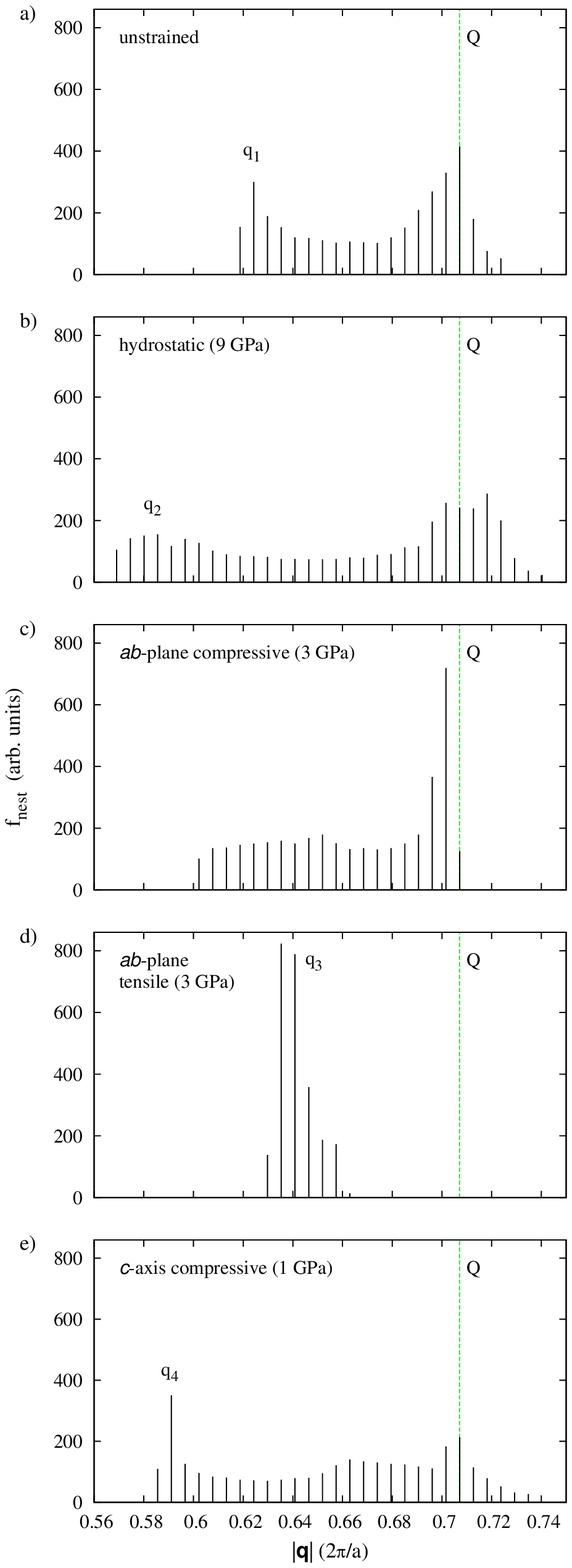}
\caption{Histograms representing the nesting function, $f_{nest}$ vs. lengths of possible spanning vectors $\mathbf{q}$ for the $\beta$ and $\delta$ FS sheets of FeSe under different kinds of strain. The length of {the ideal vector} $\mathbf{Q}$ (0.7071) is marked by vertical dashed lines.}
\label{Fig7}
\end{figure}

As visible in {fig. 7(c) and (d)}, the {{\it ab}-plane} compressive and tensile strains act otherwise than hydrostatic pressure. The {main} peaks in the $f_{nest}$ function, visible in these figure parts, {being centered} at close to $\mathbf{Q}$ (0.701) and shorter $\mathbf{q_3}$ (0.635) vectors, respectively, have both distinctly higher magnitudes than those in the other cases. Thus, especially the {onset} of the tensile strain reflects a rapid change of the nesting properties (complete lack of nesting with $\mathbf{Q}$). Since our results are limited to 3 GPa of {pressure}, which corresponds to merely 2-3\% of lattice mismatch, the FS nesting {features} of the MgO-based FeSe thin films\cite{Nie} with even larger lattice parameters are expected to be considerably different from that in the unstrained bulk FeSe. Our results, supporting the spin-fluctuation scenario of the SC pairing mechanism, are in good agreement with the experimental studies 
on a suppression or an absence of SC in the strained FeSe films \cite{Nie,Wang,Han,Li}.

Similar effects on the FS in FeSe to that of the hydrostatic pressure are obtained for the {\it c}-axis strain, as seen in {fig. 7(e)}. This influence is very strong and even relatively small values of {pressure}, e.g. 1 GPa ({fig. 7(e)}), {cause a strong} diminishing {of the intensities of nesting with the $\mathbf{Q}$ vector}. Hence, the enhancement of $T_c$ in the relative {\it c}-axis strained Fe$_{0.5}$Se$_{0.5}$ thin films\cite{Si} may also be explained by a suppression of spin ordering and an occurrence of spin fluctuations.

It is apparent that the electronic structure of FeSe plays an important role in its SC, since there is a clear dependence of the $T_c$'s values, reported in the literature for various strained FeSe systems, on the determined here FS nesting features. Thus, our findings support also the general idea of spin-fluctuation mediated SC in iron chalcogenides {\cite{Subedi, Ciechan, Singh, Imai}}.

\section{\label{sec4}Conclusions}

The influence of various strains on the structural and electronic properties of the FeSe superconductor {has} been investigated from first principles. Our results show that large modifications of the FS nesting properties in this compound can be reached by applying even a small {\it c}-axis strain. In particular, {these} should occur in thin films as well as may be induced by chemical pressure. {At the same time, the {\it ab}-plane strains} lead to opposite effects. The clear relation between the nesting features of the Fermi surface and superconducting critical temperatures, {presented in this paper, indeed confirms the earlier reports that} superconductivity in the FeSe-based systems is mediated by spin fluctuations.

\acknowledgments
This work has been supported by the EC through the FunDMS Advanced Grant
of the European Research Council (FP7 âIdeasâ) as well as by the National Center for Science in Poland (Grant No. N N202 239540 and Grant No. DEC-2011/01/B/ST3/02374). The calculations were partially performed on the ICM supercomputers of Warsaw University (Grant No. G46-13) and in Wroc\l aw Centre for Networking and Supercomputing (Project No. 158).

\end{document}